\documentclass[10pt,conference,review]{IEEEtran}
\IEEEoverridecommandlockouts

\usepackage{enumitem}
\usepackage{graphicx}
\usepackage{subfigure}
\usepackage{multirow}
\usepackage{multicol}
\usepackage{makecell}
\usepackage{xcolor}
\usepackage{amsmath,amssymb,amsfonts}
\usepackage{xcolor}
\usepackage{comment}

\renewcommand\fbox{\fcolorbox{blue}{white}}
\usepackage{balance}
\usepackage{hyperref}
\begin{document}

\title{Faster or Slower? Performance Mystery of Python Idioms Unveiled with Empirical Evidence
\vspace{-2mm}
}
\author{\IEEEauthorblockN{Zejun Zhang\IEEEauthorrefmark{1}, Zhenchang Xing\IEEEauthorrefmark{1}\IEEEauthorrefmark{3}, 
Xin Xia\IEEEauthorrefmark{2}, 
Xiwei Xu\IEEEauthorrefmark{3}, Liming Zhu\IEEEauthorrefmark{3},
Qinghua Lu\IEEEauthorrefmark{3}
}
\IEEEauthorblockA{\IEEEauthorrefmark{1}Australian National University, Australia}
\IEEEauthorblockA{\IEEEauthorrefmark{2}Software Engineering Application Technology Lab, Huawei, China}
\IEEEauthorblockA{\IEEEauthorrefmark{3}Data61, CSIRO, Australia}
\IEEEauthorblockA{\{zejun.zhang, zhenchang.xing\}@anu.edu.au,
\{xin.xia\}@acm.org, 
\{xiwei.xu, liming.zhu, qinghua.lu\}@data61.csiro.au}
}

\maketitle

\begin{abstract}
The usage of Python idioms is popular among Python developers in a formative study of 101 Python idiom performance related questions on Stack Overflow, we find that developers often get confused about the performance impact of Python idioms and use anecdotal toy code or rely on personal project experience which is often contradictory in performance outcomes. 
There has been no large-scale, systematic empirical evidence to reconcile these performance debates. 
In the paper, we create a large synthetic dataset with 24,126 pairs of non-idiomatic and functionally-equivalent idiomatic code for the nine unique Python idioms identified in~\cite{zhang2022making}, and reuse a large real-project dataset of 54,879 such code pairs provided in~\cite{zhang2022making}.
We develop a reliable performance measurement method to compare the speedup or slowdown by idiomatic code against non-idiomatic counterpart, and analyze the performance discrepancies between the synthetic and real-project code, the relationships between code features and performance changes, and the root causes of performance changes at the bytecode level.
We summarize our findings as some actionable suggestions for using Python idioms.


\end{abstract}

\section{Introduction}
Python supports many unique idioms that are designed to make the Python code concise and improve runtime performance~\cite{alexandru2018usage,zhang2022making,leelaprute2022does}. Zhang et al.~\cite{zhang2022making} recently identify nine unique Python idioms (including list/set/dictionary comprehension, chain comparison, truth value test, loop else, assignment multiple targets, star in function calls, and for multiple targets) by contrasting the language syntax of Python and Java.
We focus on these nine Python idioms in this study.
We find that developers often do not have a clear understanding of the performance impact of these Python idioms, and often present contradictory results on Stack Overflow (see Section~\ref{sec:motivation}).
Unfortunately, there has been no a large-scale systematic investigation of the performance impact of Python idioms.

A challenge to study the performance impact of Python idioms is the lack of datasets with pairs of idiomatic Python code and functionally-equivalent non-idiomatic code. 
In our formative study (see Section~\ref{sec:motivation}, we find that developers usually use synthetic toy code or rely on personal experience in real-project code to argue whether Python idiom speeds up or slows down the code. 
Both synthetic code and personal experience are anecdotal and often contradict each other. 
Synthetic code are usually simplistic~\cite{jmh,chen2020perfjit,chen2019analyzing} and cannot reflect the complexity of real-project code which may involve Python libraries and complex computations. 
Although exploring the performance impact of Python idioms in real-project code bears more practical meanings, it is very difficult to obtain pairs of idiomatic and non-idiomatic code in real projects.

Recently, Zhang et al.~\cite{zhang2022making} develop a Python idiom refactoring tool that can automatically refactor non-idiomatic Python code into nine unique Python idioms.
Built on this enabling tool, we creates a large synthetic dataset and reuses the large real-project dataset provided by~\cite{zhang2022making} to study the performance impact of nine Python idioms. 
We consider both synthetic and real-project code because it allows us to reconcile the performance debates between the two sides.

For the synthetic dataset, we consider two kinds of syntactic code features (variations of AST nodes and variable scope) and two kinds of dynamic features (data property and execution path) to construct the non-idiomatic code.
We then use the refactoring tool~\cite{zhang2022making} to refactor the non-idiomatic code to idiomatic code.
We obtain 24,126 pairs of non-idiomatic and idiomatic code for the nine Python idioms (see Table~\ref{tab:statistics_dataset}).
We manually validate the correctness of the synthetic code. 
The real-project code dataset provides 54,879 pairs of non-idiomatic code and corresponding idiomatic code for the nine Python idioms from 270 successfully configured Python projects on Github.
The authors of~\cite{zhang2022making} builds this dataset by first detecting non-idiomatic code that their tool can refactor and then refactoring it to idiomatic code.
Each code pair is accompanied with a set of test cases from the project that can execute the code before and after the refactoring.

To measure and compare the code execution time reliably, we execute a piece of code in 50 Virtual Machine (VM) invocations and collect the execution time of 35 execution (excluding the first 3 warm-up executions) in each VM invocation, following the process in previous studies~\cite{georges2007statistically,traini2021software,crape2020rigorous,laaber2019software,pyperf}.
We perform the bootstrapping with hierarchical random re-sampling and replacement on both VM invocations and execution iterations, following previous studies~\cite{davison1997bootstrap,ren2010nonparametric,traini2021software, kalibera2020quantifying}.
Our results show that all the uncertainty of the performance changes on the two datasets for each Python idiom is less than 0.05, which shows the reliability of our performance measurements.


To help developers understand the performance impact of Python idioms, we structure our study by answering the three research questions:
\textit{RQ1: What is the performance impact of Python idioms?}
\textit{RQ2: How well can code features explain the performance differences caused by Python idioms?} 
\textit{RQ3: What are the root causes of performance differences caused by Python idioms and what cause the inconsistencies between synthetic and real-project code?} 
Our results show that the impact of Python idioms on code performance vary greatly across idioms, and the impact on the synthetic code and the real-project code also differ a lot.
Although Python idioms very likely result in speedup in the synthetic code, the speedup are generally small to moderate (within 2 times).
In the real-project code, using idioms more likely results in slowdown but the difference between using or not using idioms are generally small.
We find that code features (code complexity, variable scope, data properties and execution path) can well explain the performance differences in the synthetic code, but provide limited explainability for the real-project code.
By examining the bytecode differences between non-idiomatic and idiomatic code,  we find that Python idioms usually shorten the execution time by using idiom-specialized bytecode instructions or by performing specially-design operations (e.g., truth-value-test against Python pre-defined EmptySet, swap variables, loop-else), but the effects can be diminished by library objects, overloading build-in methods, function calls and complex computations in real-project code.

The main contributions of the paper as follows:

$\bullet$ The first large-scale empirical study on the performance impact of Python idioms on synthetic and real-project code.

$\bullet$ Systematic analysis of the relationships between code features and performance differences caused by Python idioms and the performance-change root causes at the bytecode level.

$\bullet$ Systematic comparison of the performance impact of Python idioms between synthetic and real-project code.

$\bullet$ A set of actionable suggestions to use Python idioms.


\section{Formative Study}\label{sec:motivation}


To understand the opinions and arguments on the performance of Python idioms, we examine Stack Overflow questions for the nine Python idioms identified by Zhang et al.~\cite{zhang2022making}. 
We search python-tagged questions for each python idiom with the idiom name and performance-related keywords such as ``slow'', ``fast'', ``performance'', ``speed'' or ``time'' by using Stack Overflow search interface~\cite{stack_overflow}. 
For each idiom, we check the returned top-30 questions and finally collect 101 questions that discuss the performance of these nine Python idioms. Table~\ref{tab:motivation_soposts} shows some representative examples. 
\#N represents the number of questions we find for each Python idiom and blue text shows the view counts.
* indicates containing contradictory outcomes.
Our key findings are: 
\textit{Python developers are concerned with the performance of Python idioms. 
However, their evidences of performance improvement or regression are generally anecdotal based on either toy code or individual project experience. 
This leads to many controversies about if and when developers should or should not use Python idioms.}
Detailed discussions can be found in our \fbox{\href{https://github.com/idiomaticrefactoring/PerformancePythonIdioms/blob/main/performance_python_idioms_supplementary_document.pdf}{supplementary document}} (see in APPENDIX A).
\begin{table}[tbp]
\vspace{-0.1cm}
\scriptsize\caption{Python Idiom Performance Related SO Questions }
  \label{tab:motivation_soposts}
  \vspace{-0.3cm}
\centering
\begin{tabular}{|p{0.7cm}<{\centering}|p{0.10cm}<{\centering}|p{2.66in}|} 
\hline
Idiom          &\#N& Question \\ 
\hline
\begin{tabular}[c]{@{}l@{}}List-\\Compre\\hension*\end{tabular}      & 26    &\begin{tabular}[c]{@{}l@{}}\href{https://stackoverflow.com/questions/19273256/how-to-speed-up-list-comprehension}{\textbf{Question: }}How to speed up list comprehension?\\
\ \ my understanding is that \textcolor{red}{for-loop is faster than} \textcolor{red}{list comprehension}.
\\\textbf{Comments: }
In all cases I've measured the time \textcolor{red}{a list  comp}\\\textcolor{red}{rehension was always faster than a standard for loop}. (\textcolor{blue}{2k times}) 
           \end{tabular}  \\ 
\hline

\begin{tabular}[c]{@{}l@{}}Set-\\Compre\\hension*\end{tabular}       & 12    & \begin{tabular}[c]{@{}l@{}}\href{https://stackoverflow.com/questions/20496536/how-do-python-set-comprehensions-work/20497109#20497109}{\textbf{Question: }}How do python Set Comprehensions work?\\ \ \ I tried timeit for speed comparisons, there is quite some difference.\\
\textbf{Answer: } List/Dict/Set comprehensions tend to be faster than\\ anything else. (\textcolor{blue}{2k times}) 
\end{tabular}  \\ 
\hline
\begin{tabular}[c]{@{}l@{}}Dict-\\Compre\\hension*\end{tabular}      & \begin{tabular}[c]{@{}l@{}}15\end{tabular}   & \begin{tabular}[c]{@{}l@{}}\href{https://stackoverflow.com/questions/52542742/why-is-this-loop-faster-than-a-dictionary-comprehension-for-creating-a-dictionar/52542927#52542927}{\textbf{Question: }}Why is \textcolor{red}{this loop faster than a dictionary} \textcolor{red}{comprehension}?\\
\textbf{Comment:} ... I do this with a dictionary with 1000 random keys \\and values, \textcolor{red}{the dictcomp is marginally slightly} \textcolor{red}{faster}. (\textcolor{blue}{9k times}) 
\end{tabular}            \\ 
\hline
\begin{tabular}[c]{@{}l@{}}Chain-\\Compa\\rison*\end{tabular}        & \begin{tabular}[c]{@{}l@{}}6 \end{tabular}    & \begin{tabular}[c]{@{}l@{}}\href{https://stackoverflow.com/questions/34014906/is-x-y-z-faster-than-x-y-and-y-z?noredirect=1&lq=1}{\textbf{Question: }}Is "x < y < z" \textcolor{red}{faster than} "x < y and y < z"?\\
\ \ \href{https://wiki.python.org/moin/PythonSpeed}{In this page}, we know that chained comparisons are \textcolor{red}{faster than} u-\\sing the "and" operator. However, I got \textcolor{red}{a different result}... It seems \\that \texttt{x < y} and \texttt{y < z} is faster than \texttt{x < y < z}. (\textcolor{blue}{11k times})
\end{tabular}    \\ 
\hline

\begin{tabular}[c]{@{}l@{}}Truth-\\Value-\\Test* \end{tabular}       & 17    & \begin{tabular}[c]{@{}l@{}}\href{https://stackoverflow.com/questions/9850245/bool-value-of-a-list-in-python}{\textbf{Question: }}bool value of a list in Python.\\
\textbf{Answer:} \textcolor{red}{99.9\%} of the time, \textcolor{red}{performance doesn't matter} as suggest-\\ed Keith. I only mention this because I once had a scenario, using \\implicit truthiness testing \textcolor{red}{shaved 30\% off the runtime}. (\textcolor{blue}{31k times})\end{tabular} \\
\hline

\begin{tabular}[c]{@{}l@{}}Loop-\\Else\end{tabular}               & 7    & \begin{tabular}[c]{@{}l@{}}\href{https://stackoverflow.com/questions/685758/pythonic-ways-to-use-else-in-a-for-loop}{\textbf{Question: }}Pythonic ways to use 'else' in a for loop.\\
\textbf{Answer:} I was introduced to a wonderful idiom in which you can \\use a for/break/else scheme with an iterator to \textcolor{red}{save both time} and\\ LOC. (\textcolor{blue}{1k times})\end{tabular}             \\ 
\hline
\begin{tabular}[c]{@{}l@{}}Assign-\\Multi-\\Targets*\end{tabular}    & \begin{tabular}[c]{@{}l@{}}8\end{tabular}    & \begin{tabular}[c]{@{}l@{}}
\href{https://stackoverflow.com/questions/22278695/python-multiple-assignment-vs-individual-assignment-speed}{\textbf{Question: }}Python assigning two variables on one line\\
\ \ I've been looking to squeeze a little more performance out of my \\code; While browsing this \href{https://wiki.python.org/moin/PythonSpeed}{Python wiki page}, I found this claim:\\
\textcolor{red}{Multiple assignment is slower than individual assignment}.\\
I repeated several times, \textcolor{red}{but the multiple assignment snippet} \textcolor{red}{perfor-}\\\textcolor{red}{med at least 30\% better than the individual assignment}. (\textcolor{blue}{3k times}) 
\end{tabular}     \\ 
\hline
\begin{tabular}[c]{@{}l@{}}Star-in-\\Fun-Call\end{tabular} & \begin{tabular}[c]{@{}l@{}}5\end{tabular}    & \begin{tabular}[c]{@{}l@{}}\href{https://stackoverflow.com/questions/2921847/what-does-the-star-and-doublestar-operator-mean-in-a-function-call}{\textbf{Question: }}What does the star mean in a function call?\\
\ \ Does it \textcolor{red}{affect performance} at all? Is it \textcolor{red}{fast} or \textcolor{red}{slow}? (\textcolor{blue}{245k times})
\end{tabular}          \\ 
\hline
\begin{tabular}[c]{@{}l@{}}For-Mul\\-Targets\end{tabular}       & 5    & \begin{tabular}[c]{@{}l@{}}\href{https://stackoverflow.com/questions/13024416/how-come-unpacking-is-faster-than-accessing-by-index}{\textbf{\textbf{Question: }}}How come unpacking is \textcolor{red}{faster than} accessing by \\index? (\textcolor{blue}{3k times})
\end{tabular}            \\
\hline
\end{tabular}
\vspace{-0.6cm}
\end{table}
%



\section{EMPIRICAL STUDY SETUP}\label{sec:setup}
To reconcile the performance debates around Python idioms, we conduct a large-scale empirical study. 
This section describes our datasets and performance measurement method.

\subsection{Data Collection}\label{data}

Our study includes both synthetic and real-project code.
Synthetic code is generated from a set of syntactic and dynamic code features that may affect the execution time of idiomatic and non-idiomatic code.
Real-project code is from the dataset of before-after Python idiom refactorings applied to the Github projects for evaluating the Python idiom refactoring tool in~\cite{zhang2022making}.
Synthetic code provides a ``clean'' setting to measure the performance impact of Python idioms, while real-project code covers much more complex contexts in which Python idioms are used.



\subsubsection{Synthetic Dataset} 
\label{sec:syntheticdata}

\textbf{Construction Process}:
We identify a set of performance-related syntactic and dynammic code features based on the literature review~\cite{chen2020perfjit,Performance_Regression_huang,song2017performance} and our observation of Stack Overflow questions on Python idiom performance.
We vary these code features to generate diverse non-idiomatic code and then refactor it using~\cite{zhang2022making}.

\textbf{Code Features for Code Systhesis}: 
As listed in Table~\ref{tab:codefeature_lab_dataset}, syntactic features include variations of AST nodes and variable scope variation (local or global) applicable to an idiom.
Dynamic features include variant data properties and execution paths applicable to an idiom. 
The ``num*'' represents the number of AST nodes, the ``\texttt{*Set}'' represents the set of AST nodes, the ``\texttt{has*}'' (\{1 or 0\}) represents whether a Python idiom has a AST node or not, ahd the ``\texttt{is*}'' (\{1 or 0\}) represents whether a Python idiom satisfies a condition or not.

\begin{table*}[tbp]
\scriptsize
\vspace{-0.3cm}
\caption{Syntactic (AST Node and Var Scope) and Dynamic (Data Property and Execution Path) Features of Synthetic Code}
  \label{tab:codefeature_lab_dataset}
\centering
\vspace{-0.3cm}
\begin{tabular}{l| p{1.6cm}<{\centering}| p{2.2cm}<{\centering}| p{1.8cm}<{\centering}| p{1.8cm}<{\centering}| p{1.9cm}<{\centering}| c| c} 

\hline
Feature&\begin{tabular}[c]{@{}l@{}}List/Set/Dict-\\Comprehension\end{tabular}&	\begin{tabular}[c]{@{}l@{}}Chain-Comparison\end{tabular}&	Truth-Value-Test&	Loop-Else&	\begin{tabular}[c]{@{}l@{}}Assign-Multi-Targets\end{tabular}&	Star-in-Func-Call&	\begin{tabular}[c]{@{}l@{}}For-Multi-Targets\end{tabular}\\ \hline
\begin{tabular}[c]{@{}l@{}}AST Node\end{tabular} & \begin{tabular}[c]{@{}l@{}} numFor : 1-4\\
numIf : 0-4\\
numIfElse : 0-4\end{tabular} & \begin{tabular}[c]{@{}l@{}}numComop : 2-5\\ compop : \texttt{CompopSet}\end{tabular} &
\begin{tabular}[c]{@{}l@{}}test:  \texttt{TestSet} \\
compop : \texttt{EqSet}\\value : \texttt{EmptySet}\end{tabular}& \begin{tabular}[c]{@{}l@{}}\texttt{LoopSet} \\
\texttt{ConditionSet}\end{tabular}&
\begin{tabular}[c]{@{}l@{}}numAssign : 2-30\end{tabular}&
\begin{tabular}[c]{@{}l@{}}
numSubscript : 1-30
\\
\texttt{hasSubscript}\\
\texttt{hasStep}\\ 
\texttt{hasLower}\\
\texttt{hasUpper}
\end{tabular}&
\begin{tabular}[c]{@{}l@{}}
numSubscript : 1-30\\
numTarget : 1-5\\
\texttt{hasStarred}\\
\end{tabular}\\ \hline
Var Scope&\texttt{Local/Global}&\texttt{Local/Global}&\texttt{Local/Global}&\texttt{Local/Global}&\texttt{Local/Global}&\texttt{Local/Global}&\texttt{Local/Global} \\\hline
\begin{tabular}[c]{@{}l@{}}Data Prop\end{tabular}& \begin{tabular}[c]{@{}l@{}} \texttt{size} \end{tabular}& \texttt{isTrue}& \begin{tabular}[c]{@{}l@{}}\texttt{isTrue}\end{tabular}&\begin{tabular}[c]{@{}l@{}}\texttt{size}\end{tabular}& \begin{tabular}[c]{@{}l@{}}value : \texttt{isConst}\\ \texttt{isSwap} \end{tabular}&\begin{tabular}[c]{@{}l@{}}index: \texttt{isConst} \end{tabular}&\begin{tabular}[c]{@{}l@{}}\texttt{size}\end{tabular}\\\hline
Exec Path& -&-&-&\texttt{isBreak}& \begin{tabular}[c]{@{}l@{}}-\end{tabular}&-&-\\\hline
\end{tabular}
\vspace{-0.6cm}
\end{table*}

\begin{enumerate}[fullwidth,itemindent=0em,leftmargin = 0pt]
	\item[(1)] \textbf{Variations of AST Nodes:} 
Based on the Python language syntax~\cite{pythonAPI}, we construct various valid node combinations for an idiom, and if applicable, change the number of nodes to analyze the impact of these code variations. 

\noindent $\bullet$ The \textit{list/set/dict comprehension} are to append elements to an iterable. They must contain at least one \textsf{For} node, and may have \textsf{If} node with \textsf{else} or \textsf{If} node without else. We set up the range of the number of \textsf{For} node (numFor) is 1$\sim$4 (i.e., up to 4 nested loop), and the range of the number of \textsf{If} node without else (numIf) and that of \textsf{If} node with \textsf{else} (numIfElse) is 0$\sim$5. 

\noindent $\bullet$ The \textit{chain-comparison} idiom is to chain multiple comparison operations, so it has at least two comparison operators. 
We set the range of numCompop to 2$\sim$5. 
The comparison operator compop comes from the \texttt{CompopSet} (\{==, !=, $<$, $\leq$, $>$, $\geq$, is,  is-not, in, not-in\}). 

\noindent $\bullet$ The \textit{truth-value-test} idiom tests whether an object is equal or not equal to an empty value. 
The parent node of the test node belongs to \texttt{TestSet}=\{While, Assert, If\}, the comparison operator belongs to \texttt{EqSet}=\{==, !=\}, and the value belongs to Python-predefined \texttt{EmptySet}=\{None, False, `', 0, 0.0, 0j, Decimal(0), Fraction(0, 1), (), [], \{\}, dict(), set(), range(0)\}.
	
\noindent $\bullet$ The \textit{loop-else} idiom is to determine whether to execute the else clause after the iterator is exhausted. It contains one \textsf{For} or \textsf{While} node (\texttt{LoopSet}) and one \textsf{If} node with or without \textsf{else} (\texttt{ConditionSet}) after the loop block.

\noindent $\bullet$ The \textit{assign-multiple-targets} idiom is to assign multiple values in one assignment statement. 
We set the range of the number of assign node to 2$\sim$30.
For example, this idiomatic code \textsf{tar1, tar2=var1, var2} corresponds to two individual assignments in non-idiomatic code.

\noindent $\bullet$ The \textit{star-in-func-call} idiom is to unpack an iterable to the positional arguments in a function call. The non-idiomatic code consists of \textsf{Subscript} nodes (e.g., e\_list[1]). We set the range of the number of Subscript node (numSubscript) to 1$\sim$30. The corresponding idiomatic code may or may not have the Subscript node \texttt{hasSubscript}. If it has the Subscript node, it may or may not have the step, lower and upper nodes, where are denoted by \texttt{hasStep}, \texttt{hasLower} and \texttt{hasUpper}. 

\noindent $\bullet$  The \textit{for-multiple-targets} idiom is to unpack operators inside an iterable as a flattening operator. 
The non-idiomatic code has a \textsf{For} node and a \textsf{Subscript} node in the body of the \textsf{For} node to use each item. We set the range of the number of Subscript (numSubscript) node from 1$\sim$30. The corresponding idiomatic code may have many \textsf{Target} nodes, so we set the range of the number of target node to 1$\sim$5. The idiomatic code may have or not have the a \textsf{Starred} node (\texttt{hasStarred}) to represent remaining unused items. 

	\item[(2)] \textbf{Variable Scope:} In Python, the performance of variables in the local namespace is different from the variables in the global namespace~\cite{python_perf_tips,python_programming}. To explore the impact of variable scope on the code performance, we wrap the idiomatic code in a function (i.e., \texttt{Local}) or not in the function (i.e., \texttt{Global}). 
	
	\item[(3)] \textbf{Data Properties}: 
	When constructing microbenchmarks~\cite{costa2019s,jmh} to evaluate code performance, developers usually set different data sizes. 
	Inspired by this practice, we set \texttt{size} to \{$0$, $1$, $10$, $10^2$, $10^3$, $10^4$, $10^5$, $10^6$\} for an iterable. For list/set/dict comprehension, it represents the number of added elements of the iterable. For the loop-else and for-multiple-targets, it represents the number of iterations. 
	For chain-comparison and truth-value-test, we set the \texttt{isTrue} to \{0 or 1\}, representing the result of their expressions being evaluated to False or True.
	The value of assign-multiple-targets may be a temporary variable or not, which represents whether the functionality is to swap variables or not (\texttt{isSwap}=0 or 1). 
	Besides, many researchers state that constants may be faster than variables~\cite{const_perf,ismail2018quantitative,barany2014python,rodriguez2016automatic}.
	As the value of assign-multiple-targets and the index of Subscript of star-in-func-call may be a constant or not, we set \texttt{isConst}=0 or 1 for these two Python idioms. 

	\item[(4)] \textbf{Execution Path}: 
	A code may exist multiple paths, and different execution paths may lead to different performance. 
	In this work, only the loop-else involves two execution paths: whether executes the break statement or not (\texttt{isBreak}). 
\end{enumerate}


\subsubsection{Real-Project Dataset} 
We use a publicly available dataset of nine Python idiom refactorings~\cite{zhang2022making} from the 270 successfully configured GitHub Python projects.
For each instance of refactoring, it provides the successfully configured project to execute the non-idiomatic code before the refactoring and the corresponding idiomatic code after the refactoring with a set of test cases with different inputs.
To ensure that each test case executes the non-idiomatic and idiomatic code, we instrument print statements around the code to filter out the test cases without printing the content we set up. 
As both refactored code and test cases are from the real projects, we cannot control their syntactic and dynamic features.
However, we observe they still exhibit diverse feature variations.


\noindent (1) \textbf{Variations of AST Nodes}: 
For real-project code, the ``\texttt{*Set}'' is same as those for the synthetic code. 
For list-comprehension, the range of numFor, numIf, numIfElse are 1$\sim$3, 0$\sim$2 and 0$\sim$2, respectively. 
For set-comprehension, the range of numFor, numIf, numIfElse is 1$\sim$2, 0$\sim$1 and 0. 
For dict-comprehension, the range of numFor, numIf, numIfElse is 1$\sim$3, 0$\sim$1 and 0$\sim$1. 
For chain-comparison, the numCompop is 2$\sim$4. 
For assign-multiple-targets, the range of numAssign is 2$\sim$11. 
For star-in-func-call, the range of numSubscript is 2$\sim$4. 
For for-multiple-targets, the range of numSubscript and numTarget are 1$\sim$6 and 1$\sim$3, respectively. 
For star-in-func-call, 1\%, 4\%, 16\% and 16\% of code instances have \texttt{hasStep}, \texttt{hasLower}, \texttt{hasUpper} and \texttt{hasSubscript} being 1. For for-multi-targets, 90\% of code instances have \texttt{hasStarred} being 1. The code-feature variations in the real-project dataset is smaller than those in the synthetic code.

\noindent (2) \textbf{Variable Scope}: Since the real-project code is executed by test cases, all variables of Python code are in the local scope.

\noindent (3) \textbf{Data Properties}: 
For list/set/dict-comprehension, loop-else and for-multiple-targets, their ranges of \texttt{size} are 0$\sim$11,766, 0$\sim$44, 0$\sim$1,000, 0$\sim$50 and 0$\sim$583, respectively. For the chain-comparison,  45\% of \texttt{isTrue} is 1. For the truth-value-test,  25\% of \texttt{isTrue} is 1. 
For assign-multi-targets, 0.04\% of \texttt{isSwap} of code instances is 1 (i.e., rarely used for swapping variables). 
For assign-multi-targets and star-in-func-call, 1\% and 97\% of code instances have \texttt{isConst} being 1.

\noindent (4) \textbf{Execution Path}: 
For the loop-else, the percentage of code instances with \texttt{isBreak}=1 is 89\%.

\subsubsection{Dataset Summary}
Table~\ref{tab:statistics_dataset} summarizes the number of code pairs (non-idiomatic versus idiomatic) in the synthetic dataset and the real-project dataset. 
For the synthetic code, the number of code instances is obtained by multiplying the feature variants applicable to a idiom.
For example, for synthetic list/set/dict-comprehension code pairs, 1600 is computed by multiplying 4 numFor, 5 numIf, 5 numIfElse, 2 and 8 values (local or global) of variable scope and \texttt{size}. 
The calculation of the number of code pairs for other Python idioms can be found in our \fbox{\href{https://github.com/idiomaticrefactoring/PerformancePythonIdioms/blob/main/performance_python_idioms_supplementary_document.pdf}{supplementary document}} (see in APPENDIX B).
For real-project code, the number of code pairs are reported by the refactoring tool~\cite{zhang2022making} (accompanied by at least one test case to execute the code before- and after-refactoring).

\begin{table}[tbp]
\scriptsize
\vspace{-0.4cm}
\caption{The statistics of Synthetic and Real-Project Dataset }
  \label{tab:statistics_dataset}
\centering
\vspace{-0.3cm}
\begin{tabular}{l| c| c} 
 \hline
Idiom& Synthetic  & Real-Project\\
 \hline

List Comprehension  & 1600 & 734   \\ 
\hline
Set Comprehension  & 1600 &  282  \\ 
\hline
Dict Comprehension  & 1600   & 194    \\ 
\hline

Chain Comparison   & 11968  & 2268 \\ 
\hline
Truth Value Test   &336   & 40116\\ 
\hline
Loop Else    & 128& 198 \\ 
\hline
Assign Multiple Targets  & 174  & 10583 \\ 
\hline
Star in Func Call  & 1920   & 170 \\ 
\hline
For Multiple Targets & 4800 & 334 \\
\hline
Total & 24126 & 54879 \\
\hline
\end{tabular}
\vspace{-0.75cm}
\end{table}


\subsection{Performance Measurement}\label{measurement}
The measurement of execution time of non-idiomatic and idiomatic code is far from trivial due to the non-determinism such as Python Virtual Machine (VM) and garbage collector~\cite{georges2007statistically}. 
To overcome such non-determinism, researchers repeatedly execute the code multiple times~\cite{georges2007statistically,chen2020perfjit,traini2021software,ding2020towards,stefan2017unit,laaber2018evaluation,laaber2019software,crape2020rigorous}. 
Since different VM invocations may result in different code execution time, and multiple executions of the code in a VM invocation may vary, we execute each code 35 iterations on 50 VM invocations as done in previous studies~\cite{georges2007statistically,traini2021software,crape2020rigorous,laaber2019software}. 
Since first iterations (i.e., warm-up iterations) are subject to noise caused by library loading, measurements are only collected in iterations that are subsequent to warm-up. 
We set the warm-up iterations to three which is enough to warm up the benchmarks~\cite{traini2021software,pyperf}. 
We run the two datasets on the CPython interpreter version 3.7.12 on an Ubuntu 18.04.4 System.

Following previous studies surveyed in Kalibera et al.~\cite{kalibera2013rigorous,kalibera2020quantifying},
to indicate how much speedup or slowdown idiomatic code achieves compared with non-idiomatic code, we divide the total of execution time of non-idiomatic code by that of corresponding idiomatic code. 
The formula as follows: $\rho=\sum_{i=1}^{n}\sum_{j=1}^{k}m_{i,j}^{nonidiomatic}/\sum_{i=1}^{n}\sum_{j=1}^{k}m_{i,j}^{idiomatic}$
where $n$ is the number of VM invocations, $m$ is the number of measurement iterations, and $m_{i,j}$ is the execution time of the $k$th iteration in the $i$th invocation. 
The longer execution time means the slower speed.
If $\rho$ is larger than 1, the idiomatic code has a $\rho$X speedup (i.e., the idiomatic code is 1-1$/\rho$\% faster than non-idiomatic code). 
For example, if the time of non-idiomatic code is 2s and the time of idiomatic code is 1s, the idiomatic code has a 2X speedup and is 50\% faster.
If $\rho$ is less than 1, the idiomatic code has a $1/\rho$X slowdown (i.e., the idiomatic code is $1/\rho-1$ times slower than non-idiomatic code). 
If $\rho$=1, the performance of idiomatic code is the same as that of non-idiomatic code. 

To validate the reliability of performance measurement, we apply the approach of quantifying performance changes with effect size proposed by Kalibera et al.~\cite{kalibera2013rigorous,kalibera2020quantifying}. 
We use bootstrapping with hierarchical random re-sampling and replacement~\cite{davison1997bootstrap,ren2010nonparametric,traini2021software} on two levels~\cite{traini2021software,kalibera2020quantifying}: VM invocations and iterations.
We run the experiments 1000 times and obtain 1000 performance change $P = \{\rho_i \mid 1 \le i \le 1000\}$.
We obtain the lower limit $\rho_l$ and upper limit $\rho_u$ for 95\% confidence from $P$. Then we compute uncertainty of performance change by the $(\rho_u-\rho_l)/(\rho)$. 
Our results show that all the uncertainty of the performance change on the two datasets for each Python idiom is less than 0.05, which shows the reliability of our performance measurements. 

\section{EMPIRICAL ANALYSIS}\label{sec:result}
\subsection{RQ1: What is the performance impact of python idioms? Is the impact consistent on synthetic and real-project code?}

\subsubsection{Motivation}

Section~\ref{sec:motivation} shows that developers are concerned with the performance of Python idioms and often experience different and even contradictory results.
However, there lack of large-scale, systematic empirical evidence of the performance differences between functionally-equivalent idiomatic and non-idiomatic code.
Furthermore, there is little consensus between the performance impact on synthetic and real-project code.
This RQ fills in this gap.

\subsubsection{Approach}
We consider nine unique Python idioms~\cite{zhang2022making}.
As described in Section~\ref{data}, we synthesize a large dataset of 24,126 pairs of functionally-equivalent idiomatic and non-idiomatic code with variant syntactic and dynamic code features.
We also use a large dataset of 54,879 pairs of functionally-equivalent idiomatic and non-idiomatic code collected by~\cite{zhang2022making}.
We adopt a systematic and reliable execution time benchmarking method to measure the performance differences between a pair of code. 
We analyze the performance differences from four aspects: 
performance impact (speedup, slowdown or unchanged), maximum speedup, maximum slowdown, and the variation of performance changes.  

\subsubsection{Results}

Fig.~\ref{fig:performance_rq1} shows the distribution of performance differences between non-idiomatic and idiomatic code of nine Python idioms on the two datasets. 
The gray circles are the outliers.
The orange line inside the box represents the median value. 
The upper and lower whiskers represent the maximum speedup and slowdown excluding the outliers, respectively. 
The box represents the 25th to 75th percentile of dataset.

\noindent$\bullet$ \textbf{Performance speedup, slowdown or unchanged:} 
On the synthetic code, the majority of $\rho$ for seven idioms is $>$1, including list/set/dict-comprehension, truth-value-test, loop-else, star-in-func-call, and for-multi-targets.
That is, for these seven idioms, idiomatic code more likely results in performance speed up against non-idiomatic code.
Only one idiom (i.e., assign-multi-targets) more likely results in performance slowdown (i.e., the majority of $1/\rho>$1 or $\rho<$1). 
Except for loop-else, no matter the majority is speedup (or slowdown), there are always certain percentage of cases having the opposite effect.
For chained-comparison, about half of the cases result in speedup, while the other half result in slowdown.

On the real-project code, six idioms most likely result in performance slowdown (i.e., the majority of $1/\rho>$1), including list-comprehension, dict-comprehension, chain-comparison, assign-multi-targets, star-in-func-call, and for-multi-targets.
Only one idiom (truth-value-test) more likely results in performance speedup.
For the two idioms (set-comprehension, loop-else), the majority of cases with $\rho$ very close to 1 (except for some outliers). 
We consider the performance difference of these two idioms as unchanged.

Across the two datasets, only truth-value-test and assign-multi-targets exhibit consistent performance impact (the majority speedup or slowdown).
Four idioms exhibit the opposite impact (the majority speedup vs. the majority slowdown).
The other three idioms also exhibit different impacts (the majority speedup vs. the majority unchanged for set-comprehension and loop-else, and half-half split vs. the majority slowdown for chained-comparison).

\noindent$\bullet$ \textbf{Maximum speedup:} 
On the synthetic code, six idioms (list/set/dict-comprehension, truth-value-test, star-in-func-call, for-multi-targets) have the maximum speedup $>$2.
Three idioms (list-comprehension, truth-value-test and for-multi-targets) have relatively more percentages (15\%, 15\% and 19\% respectively) of cases with $>$2 speedup.
Three idioms (set-comprehension, dict-comprehension and star-in-func-call) have a small number of cases (0.3\%-5\%) with $>$2 speedup.
Three idioms (chain-comparison, loop-else, assign-multi-targets) have the maximum speedup $<$2.
The speedup of some outliers for list-comprehension, set-comprehension, truth-value-test and for-multi-targets can be $>$4 times.
On real-project code, although only truth-value-test has the majority speedup, it has 28\% of cases with $>$2 speedup. 
The maximum speedup for truth-value-test reaches about 11 times.
All other eight idioms have the maximum speedup $<$2 (even the outliers are not close to 2).
Among these eight idioms, except for list-comprehension, star-in-fun-call and for-multi-targets, five idioms have the maximum speedup close to 1 (i.e., the same as non-idiomatic code).
With the 2 times speedup as a threshold, the maximum speedup of truth-value-test, chained-comparison, loop-else and assign-multi-targets are consistent across the two datasets (both $>$2 or both $<$2).
The maximum speedup of the other five idioms are inconsistent.

\noindent$\bullet$ \textbf{Maximum slowdown:} 
On the synthetic code, only for-multi-targets has the maximum slowdown $>$2, accounting for 15\% of cases.
The slowdown of some for-multi-targets outliers reaches about 8 times.
For the other eight idioms, the maximum slowdown is all $<$2, except for a few outliers.
Among these eight idioms, truth-value-test and loop-else have the maximum slowdown close to 1.
On the real-project code, truth-value-test and for-multi-targets have the maximum slowdown $>$2, accounting for 10\% and 27\% of the cases of each idiom, respectively.
An outlier of for-multi-targets reaches 6 times slowdown, and an outlier of truth-value-test reaches 20 times slowdown.
For the other seven idioms, the maximum slowdown is all $<$2, except for a few outliers.
Loop-else has the maximum slowdown close to 1.
With the 2 times slowdown as a threshold, the maximum slowdown of eight idioms are consistent (both $>$2 or both $<$2).
Only only one idiom (truth-value-test) has inconsistent maximum slowdown.

\noindent$\bullet$ \textbf{The variation of performance changes:}
We calculate the variation between the maximum speedup and the maximum slowdown.
On the synthetic code, two idioms (list-comprehension and for-multi-targets) have the variation $>2$.
Four idioms (set/dict-comprehension, truth-value-test, star-in-func-call) have the variation between 1 and 2.
The other three idioms (chained-comparison, loop-else, assign-multi-targets) have the variations $<$1.
On the real-project code, the variation of performance changes of truth-value-test and for-multi-targets is $>$ 2. 
For the other seven idioms, the variation of performance changes is $<$ 1.
The performance changes of three idioms (chained-comparison, loop-else, assign-multi-targets) are relatively stable on both datasets, and the performance changes of for-multi-targets vary largely on both datasets.
Except for truth-value-testing, the variation of 8 idioms on real-project code is smaller than that on synthetic code.
\noindent\fcolorbox{red}{white}{\begin{minipage}{8.6cm} \emph{The impact of Python idioms on code performance vary greatly across idioms and between synthetic code and real-project code. Python idioms do not always result in performance speedup. In general, most idioms (except for assign-multi-targets) more likely result in speedup in the synthetic code, while most idioms (except for truth-value-test and loop-else) more likely result in slowdown in the real-project code. The slowdown that an idiom causes is usually within 2 times, while the speedup can sometimes be above 2 times. List-comprehension, truth-value-test and for-multi-targets may cause significant speedup, but caution should be applied when using truth-value-test and for-multi-targets as they may also cause significant slowdown.} \end{minipage}}

\subsection{RQ2: How well can code features explain the performance differences caused by Python idioms?}\label{rq2}
\subsubsection{Motivation}

Based on the literature review~\cite{alexandru2018usage,phan2020teddy,zhang2022making,programming_idioms} and the observation of the performance-related Stack Overflow questions, we identify a set of syntactic and dynamic features (see Section~\ref{sec:syntheticdata}) that may potentially affect code performance.
However, it is unclear whether and how some or all these code features actually correlate with the performance differences caused by Python idioms.
If the correlations exist, making them explicit would help developers determine when to use Python idioms and anticipate their effects.

\begin{figure}[tbp]
  \centering
  \includegraphics[width=2.9in]{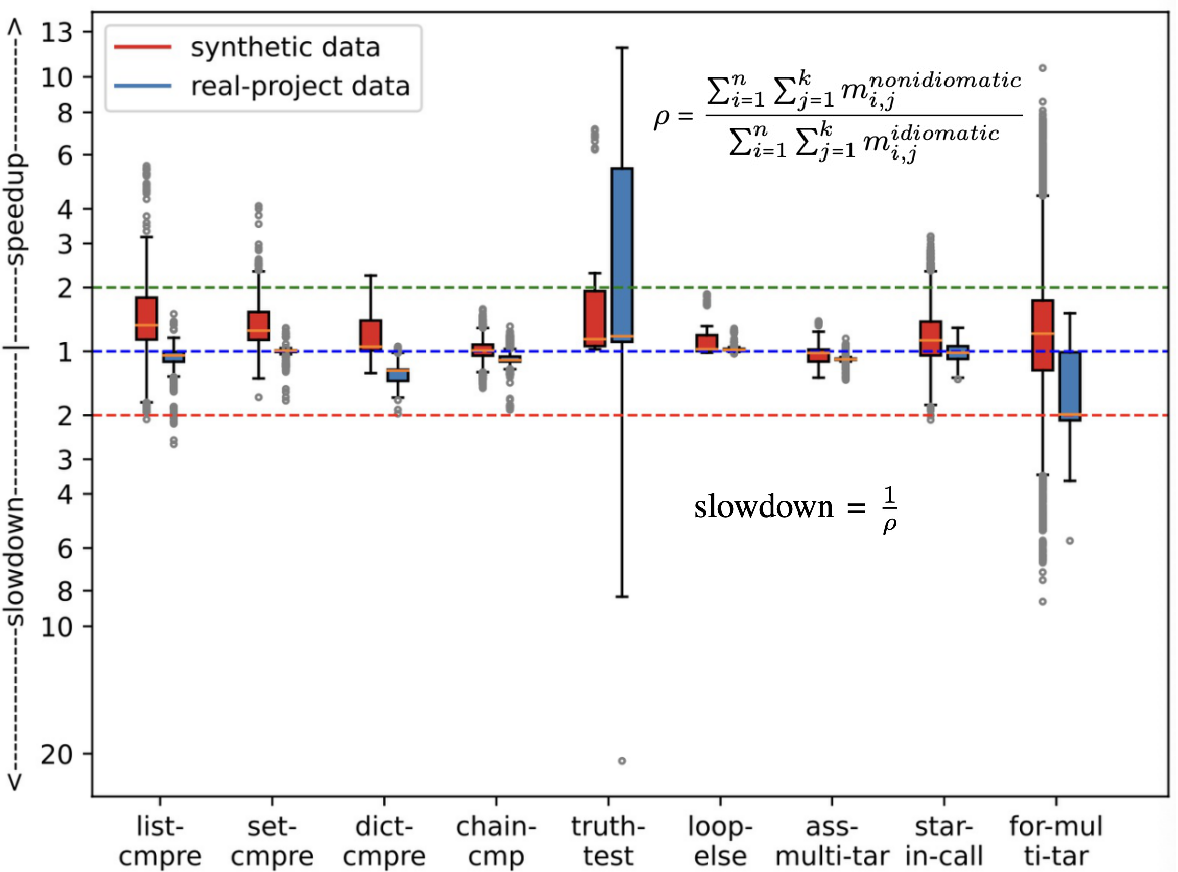}
\setlength{\abovecaptionskip}{-0.1cm}
    \caption{Performance Differences between Idiomatic vs. Non-idiomatic Code}
    
    \label{fig:performance_rq1} 
    \vspace{-0.7cm}
\end{figure}

\subsubsection{Approach}

To analyze the relationships between the code features and the performance differences reported in RQ1, we construct a Generalized additive model (GAM)~\cite{tan2022exploratory,hastie2017generalized} using by the package ``mgcv'' in R language. 
GAM can analyze complex nonlinear and non-monotonic relationships between the code features and the performance differences~\cite{hastie2017generalized}. 
It also provides good interpretability.
Similar to previous studies~\cite{hastie2017generalized,tan2022exploratory}, we log-transform highly skewed numeric code features and performance changes. 
We use the variance inflation factor (VIF) to analyze the multicollinearity of the two features~\cite{mansfield1982detecting}. 
If the VIF of two features is larger than 5~\cite{JSSv032b01,tan2022exploratory,articleVIFZagreb}, there is significant multicollinearity between the two features, we randomly remove one of them. 

The GAM model provides deviance explained (DE), p-values and the trend plot between the code features and performance differences which help us understand the positive or negative correlations. 
The DE value indicates the goodness-of-fit of model. 
The closer the DE is to 1, the better the model fits the data. 
Previous studies generally assume that the model can explain the data well if the DE value is greater than 40\%~\cite{tan2022exploratory,SCHOEMAN200267,articleVIFZagreb}. 
The p-value for each feature tests the null hypothesis that the feature has no correlation with the performance difference. 
If the p-value for a feature is less than 0.05, the feature is regarded as important~\cite{tan2022exploratory,SCHOEMAN200267}. 
Similar to the previous study~\cite{GAM30days}, we sort the features by the change of DE after removing one important feature. 
The large the change, the more important the feature.

\subsubsection{Result}

\begin{table}[tbp]
\vspace{-0.3cm}
\scriptsize\caption{The Correlations between Code Features and Performance Differences Caused by Python Idioms.}
  \label{tab:relation_code_performance}
\centering
\vspace{-0.3cm}
\begin{tabular}{l|c|l|c|l}
\hline
\multirow{2}{*}{Idiom}                                    & \multicolumn{2}{c|}{Synthetic Code} & \multicolumn{2}{c}{Real-Project Code}\\
\cline{2-5} &\begin{tabular}[c]{@{}l@{}}DE (\%) \end{tabular} & \multicolumn{1}{l|}{\begin{tabular}[c]{@{}l@{}}Important Features\end{tabular}}&\multicolumn{1}{c|}{\begin{tabular}[c]{@{}l@{}}DE (\%) \end{tabular}} &\multicolumn{1}{l}{\begin{tabular}[c]{@{}l@{}}Important Features\end{tabular}}\\\hline
\begin{tabular}[c]{@{}l@{}}List-Com\\prehension\end{tabular}  &  90.3& \multirow{3}{*}{\begin{tabular}[c|]{@{}l@{}}
\textcolor{blue}{\texttt{size}}\\
\textcolor{blue}{\texttt{scope}=Global}\\
 \textcolor{red}{numFor}\\
  \textcolor{red}{numIf}\\
    \textcolor{red}{numIfElse}\\
\end{tabular}} & \multicolumn{1}{|c|}{28.3} & \multirow{3}{*}{-} \\ \cline{1-2} \cline{4-4}
\begin{tabular}[c]{@{}l@{}}Set-Comp\\rehension\end{tabular}                                            & 88.1 &                   & \multicolumn{1}{|c|}{8.8} &                   \\ \cline{1-2} \cline{4-4}
\begin{tabular}[c]{@{}l@{}}Dict-Comp\\rehension\end{tabular}                                            & 91.2 &                   & \multicolumn{1}{|c|}{29.3} &                   \\ \cline{1-1}
\hline

\begin{tabular}[c]{@{}l@{}}Chain\\Comparison\end{tabular}   & 55.4&\multicolumn{1}{l|}{\begin{tabular}[c]{@{}l@{}}
\textcolor{blue}{\texttt{scope}=Global}\\
\textcolor{blue}{numComop}\\
\end{tabular}}   & \multicolumn{1}{c|}{19.8} &\begin{tabular}[c]{@{}l@{}}-
\end{tabular}\\ 
\hline
\begin{tabular}[c]{@{}l@{}}Truth\\Value\\Test\end{tabular}   & 99.6&\multicolumn{1}{l|}{\begin{tabular}[c]{@{}l@{}}
\textcolor{blue}{\texttt{EmptySet}}\\
\textcolor{red}{\texttt{scope}=Global}\end{tabular}}   &\multicolumn{1}{c|}{32}&\begin{tabular}[c]{@{}l@{}}-
\end{tabular}\\ 
\hline
\begin{tabular}[c]{@{}l@{}}Loop\\Else\end{tabular}    & 95.1&\multicolumn{1}{l|}{\begin{tabular}[c]{@{}l@{}}
 \textcolor{red}{\texttt{size}}\\
 \textcolor{blue}{\texttt{isBreak}=1}\end{tabular}} & \multicolumn{1}{c|}{88.7} & \multicolumn{1}{l}{\begin{tabular}[c]{@{}l@{}}\textcolor{red}{\texttt{size}}\\
  \textcolor{blue}{\texttt{isBreak}=1}
 \end{tabular}}\\ 
\hline
\begin{tabular}[c]{@{}l@{}}Assign\\Multi\\Targets\end{tabular}    & 68.6 &\multicolumn{1}{l|}{\begin{tabular}[c]{@{}l@{}}
 \textcolor{blue}{\texttt{isSwap}=1}\\ \textcolor{blue}{\texttt{scope}=Global}\\
 \textcolor{blue}{\texttt{isConst}=1}
\end{tabular}} & \multicolumn{1}{c|}{55.1}& \multicolumn{1}{l}{\begin{tabular}[c]{@{}l@{}}
\textcolor{red}{numAssign}\\
\textcolor{blue}{\texttt{isConst}=1}
\end{tabular}}\\ 
\hline
\begin{tabular}[c]{@{}l@{}}Star-in\\Func-Call\end{tabular}  & 90.9&\multicolumn{1}{c|}{\begin{tabular}[c]{@{}l@{}}
 \textcolor{blue}{\texttt{numSubscript}}\\
 \textcolor{blue}{\texttt{scope}=Global}\\ 
  \textcolor{red}{\texttt{isConst}=1}\\
  \textcolor{red}{\texttt{hasSubscript}=1}\\
\textcolor{red}{\texttt{hasStep}=1}\\
\textcolor{red}{\texttt{hasUpper}=1}\\
\textcolor{red}{\texttt{hasLower}=1}
\end{tabular}}   & \multicolumn{1}{c|}{31.9} &\begin{tabular}[c]{@{}l@{}}-\end{tabular}\\ 
\hline
\begin{tabular}[c]{@{}l@{}}For\\Multi\\Targets\end{tabular} & 88.3&\multicolumn{1}{l|}{\begin{tabular}[c]{@{}l@{}}
 \textcolor{blue}{\texttt{numScript}}\\
   \textcolor{red}{\texttt{hasStarred}=1}\\ \textcolor{red}{\texttt{scope}=Global}\\
   \textcolor{red}{numTarget}\\
\textcolor{blue}{\texttt{size}}
\end{tabular}}  & \multicolumn{1}{c|}{72.8}&\multicolumn{1}{l}{\begin{tabular}[c]{@{}l@{}} 
 \textcolor{blue}{\texttt{numScript}}\\
   \textcolor{red}{\texttt{hasStarred}=1}\\ 
   \textcolor{red}{numTarget}\\
   \textcolor{blue}{\texttt{size}}
\end{tabular}}\\
\hline
\end{tabular}

\textbf{Note}: \textcolor{blue}{Blue} - Positive Correlation, \textcolor{red}{Red} - Negative Correlation
\vspace{-0.8cm}
\end{table}

Table~\ref{tab:relation_code_performance} presents our analysis results.
For the synthetic code, the DE values for all nine idioms are above 40\%.
That is, code features can explain the performance differences caused by all nine idioms in the synthetic code.
We list the important features ($p$-value$<$0.05) for each idiom in the descending order of importance.
In contrast, only three idioms (loop-else, assign-multi-targets, for-multi-targets) have the DE values above 40\% for real-project code.
That is, except for these three idioms, code features alone cannot explain the performance differences caused by the other six idioms in the real-project code.
For these six idioms, as feature importance becomes meaningless, their important feature cells show ``-''.
If a feature change (value increase for num*, setting 1 for is* and has*, using a specific value in a set) incurs the larger performance speedup or smaller slowdown, we say the feature is positively correlated to the performance change (shown in blue font).
Otherwise, the feature is negatively correlated to the performance change (shown in red font).

\noindent$\bullet$ \textbf{Analysis of Synthetic Code:}
For the synthetic code, code complexity features (num* and has*) generally negatively correlate with the performance changes, while data size generally positively correlate with the performance changes.
Code features specially-designed for idioms (e.g., swapping variable, assigning constant value, empty value testing) positively correlate with the performance speedup.
Variable scope (local versus global) has both ways of correlations.

\textit{Code complexity:}
When list/set/dict-comprehension involves deeper nested loops (numFor increases) and more if-condition checking (numIf and numIfElse increases), the performance speedup by using list/set/dict-comprehension becomes smaller.
For star-in-func-call, when using a subscript (hasSubscript=1), the more complex the subscript becomes (using step, lower, upper), the smaller the performance improvement becomes.
For for-multi-targets, accessing more targets (numTarget increases) makes the performance speedup smaller, and using a starred node (hasStarred=1) makes slowdown even larger.
There are several exceptions to this general negative correlations, including numCompop for chained-comparison, numSubscript for star-in-func-call and for-multi-targets.
That is, the increase of comparisons and subscripts result in more speedup.
We believe this is reasonable as these three idioms are specially designed to handle more comparisons and more subscripts efficiently.

\begin{figure}[tbp]

\centering

\begin{minipage}{2.9cm}
\centering
\vspace{-0.1cm}
\includegraphics[width=2.9cm]{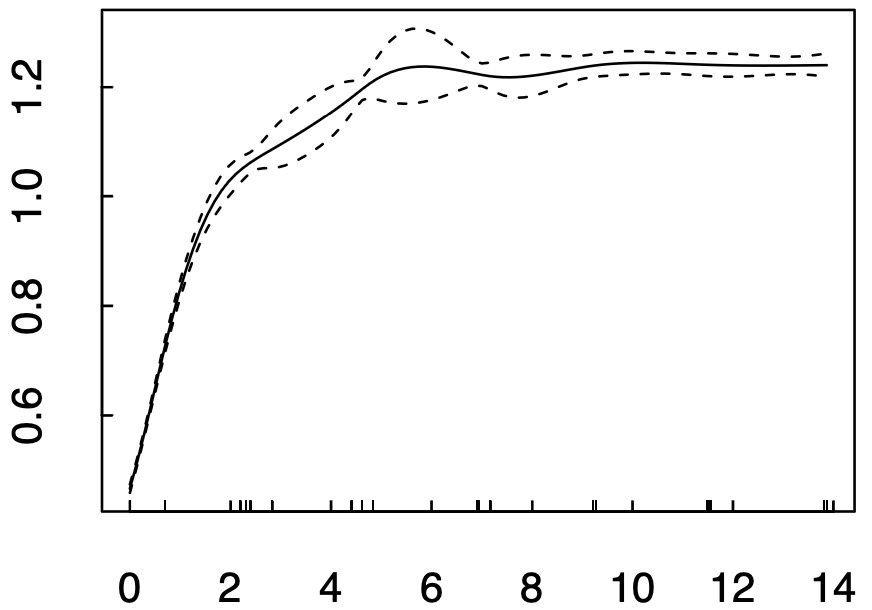}
\vspace{-0.1cm}
\end{minipage}%
\begin{minipage}{2.9cm}
\centering
\vspace{-0.1cm}
\includegraphics[width=2.9cm]{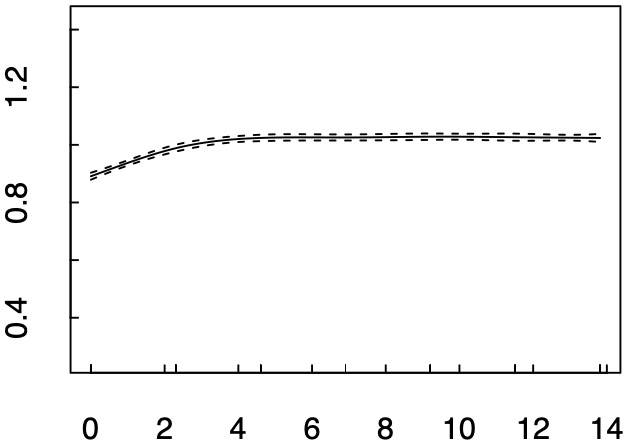}
\vspace{-0.1cm}
\end{minipage}
\begin{minipage}{2.9cm}
\centering
\vspace{-0.1cm}
\includegraphics[width=2.9cm]{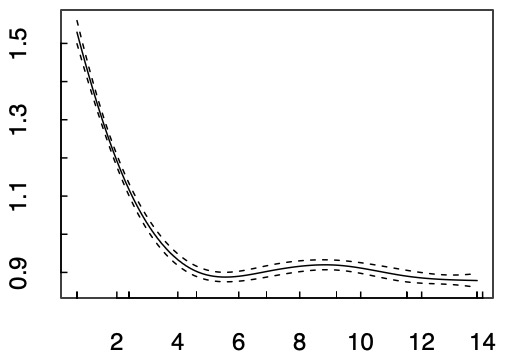}
\vspace{-0.1cm}

\end{minipage}

\caption{The relationship between \texttt{size} and performance changes for list comprehension, for-multi-targets and loop-else}
\vspace{-0.8cm}
\label{fig:sizeperformance} 

\end{figure}

\textit{Data size:}
The more elements for list/set/dict-comprehension and the more iterations for for-multi-targets (data size increases), the more performance speedup these idioms produce.
However, this speedup stops or increases only marginally once the data size reaches a certain limit. 
Fig.~\ref{fig:sizeperformance} shows this observation for list-comprehension and for-multi-targets. 
To see the trend clearly, we take the log of data size.
The figures for set/dict comprehension can be found in our \fbox{\href{https://github.com/idiomaticrefactoring/PerformancePythonIdioms/blob/main/performance_python_idioms_supplementary_document.pdf}{supplementary document}} (see APPENDIX C.A).
For example, for list-comprehension, the speedup increases fast when the number of elements increases from 1 to 55 ($e^{4}$).
However, the speedup flattens when the number of elements increases over about 2981 ($e^{8}$).
The impact of data size on for-multi-targets is smaller than that on list/set/dict-comprehension.
In contrast, the data size for loop-size is negatively correlated with the speedup.
The more iterations the loop runs, the smaller speedup the loop-else idiom produces.
This is because the speedup by the special else processing accounts for smaller and smaller percentage of the overall execution as the loop runs more iterations.

\textit{Specific code features:}
Truth-value-test is specially designed to handle the Python pre-defined empty values in the EmptySet. 
So some specific empty values may produce significant speedup.
For example, truth-value-testing ``Fraction(0,1)'' has 6 times speedup against non-idiomatic code.
For other values in the EmptySet, the speedup ranges from 1.02 to 2.3 times. 
When assign-multi-targets is used to swap variables or to assign constant values, the idiom produces smaller slowdown than processing regular multiple variable assignments. 
When using the constant subscript index in star-in-func-call, the speed up is smaller than using non-constant index.

\textit{Variable scope:}
The impact of variable scope varies on different idioms.
For list/set/dict-comprehension and chain-comparison, using global scope results in larger speedup, while for truth-value-test, assign-multi-targets, star-in-func-call and for-multi-targets, the effect is the opposite.
Variable scope is not an important feature affecting loop-else.

\textit{Execution path:} For the loop-else, \texttt{isBreak} positively correlate with the performance changes. Since executing the break statement means the loop run fewer iterations than not executing the break statement, the speedup when \texttt{isBreak}=1 is greater than when \texttt{isBreak}=0.


\noindent$\bullet$ \textbf{Analysis of Real-Project Code:}
For loop-else on real-project code, it important features (\texttt{size} and \texttt{isBreak=1}) and the causes of correlation are the same as loop-else on the synthetic code (seen the above discussion on data size and execution path).
For assign-multi-targets on real-project code, numAssign and \texttt{isConst} are important features, which overlap only one feature with the three important features (\texttt{isSwap}, scope, \texttt{isConst}) for the idiom on synthetic code.
The positive correlation of \texttt{isConst} is the same on the synthetic and real-project code.
Scope variant is irrelevant to real-project code as it is always local scope.
\texttt{isSwap} is not applicable to real-project code as assign-multi-targets has rarely been used to swap variables in the real-project code.
For for-multi-targets, the set of important features (numSubscript, numTarget, \texttt{size}, \texttt{hasStarred}) and the impact of these features are the same on the synthetic and real-project code. 
Again, the scope variant is irrelevant to real-project code. 

For the other six idioms (list/set/dict-comprehension, chained-comparison, truth-value-test, star-in-func-call), none of the code features have important correlations with the performance changes caused these idioms.
As further studied in Section~\ref{sec:rootcause}, many external factors (e.g., using library objects) may affect the performance change in the real-project code, other than the code features studied in the RQ2.

\vspace{0.1cm}
\noindent\fcolorbox{red}{white}{\begin{minipage}{8.6cm} \emph{Code complexity, data size, execution path and specific code features can well explain the performance differences caused by Python idioms on the synthetic code, but offer only limited explainability on the real-project code. Only loop-else and for-multi-targets share the same important features on the two types of code. In general, higher code complexity diminishes the effect of Python idioms. 
Python idioms benefit large data size (with upper bound though) and specific code functions they are designed for. However, these benefits may be diminished by the complex factors in real-project code.} \end{minipage}}

\subsection{RQ3: What are the root causes of performance differences caused by Python idioms and what cause the inconsistencies between synthetic and real-project code?}
\label{sec:rootcause}

\subsubsection{Motivation}

RQ1 and RQ2 show that Python idioms have complex impact on code performance and the impact is not consistent between synthetic and real-project code.
Furthermore, source code features are not sufficient to explain the performance changes in real-project code.
In this RQ, we investigate how Python idioms are implemented at the bytecode level and identify the bytecode instruction differences between pairs of non-idiomatic and idiomatic code in our datasets.
Analyzing these instruction differences allows us to explain the root causes of performance differences and the inconsistencies between synthetic and real-project code.

\subsubsection{Approach}
We get the bytecode instructions of idiomatic code and the corresponding non-idiomatic code for each Python idiom with dis module~\cite{dis_python}.
Then we compare the difference in bytecode instructions in a semi-automatic manner using code differencing tool and manual examination.
For each Python idiom in the two datasets, we analyze all code pairs if the number of code pairs is less than 400.
Otherwise, to make manual examination feasible, we randomly sample code pairs with a confidence level of 95\% and an error margin 5\%~\cite{singh2013elements}. Two authors first independently analyze the instruction differences between non-idiomatic and idiomatic code and summarize the differences into categories.
Then, they discuss and reach the consensus on the final categories.

\subsubsection{Result} 
Table~\ref{tab:statis_reason} lists the five categories of root causes.
\#N represents the number of examined code instances. 
\begin{table}[tbp]
\scriptsize
\caption{The Statistics of Root Causes of Performance Differences} 
  \label{tab:statis_reason}
  \vspace{-0.27cm}
\centering
\begin{tabular}{p{0.48cm}|p{0.2cm}<{\centering}p{0.3cm}<{\centering}p{0.3cm}<{\centering}p{0.24cm}<{\centering}p{0.23cm}<{\centering}p{0.25cm}<{\centering}p{0.3cm}<{\centering}p{0.3cm}<{\centering}p{0.3cm}<{\centering}p{0.34cm}}
\hline
\multirow{2}{*}{Idiom} & \multicolumn{4}{c|}{Synthetic Code} & \multicolumn{6}{c}{Real-Project Code} \\ \cline{2-11} 
                  & \multicolumn{1}{c|}{\#N}      & R1 & R2  & \multicolumn{1}{c|}{R3} &  \multicolumn{1}{c|}{\#N} & R1    & R2   & R3   & R4   & R5  \\ \hline
\begin{tabular}[c]{@{}l@{}}list-\\cmpre\end{tabular}  &     \multicolumn{1}{c|}{310} & 18.7   &  81.3   & \multicolumn{1}{c|}{-} &  \multicolumn{1}{c|}{253} &  15.8			   & 5.1        & -     &10.7      &  68.4     \\ \hline
\begin{tabular}[c]{@{}l@{}}set-\\cmpre\end{tabular}   &   \multicolumn{1}{c|}{310}  &  14.8  &  85.2    & \multicolumn{1}{c|}{-} &  \multicolumn{1}{c|}{\textcolor{black}{282}} &   16.1		    & 3.1   &  -    &   32.3   &   48.5  \\ \hline
\begin{tabular}[c]{@{}l@{}}dict-\\cmpre\end{tabular} & \multicolumn{1}{c|}{310}    &   20.3 & 79.7     &   \multicolumn{1}{c|}{-}  &\multicolumn{1}{c|}{\textcolor{black}{194}}                      & 38.2      &  4.6        &  -    & 3.6     & 53.6     \\\hline
\begin{tabular}[c]{@{}l@{}}chain\\-comp\end{tabular}  & \multicolumn{1}{c|}{373}         &  -    & 100   &  \multicolumn{1}{c|}{-}    &      \multicolumn{1}{c|}{329}               &  -
        & 64.7     &  -    &  -    &  35.3    \\\hline
\begin{tabular}[c]{@{}l@{}}truth-\\test\end{tabular}   &  \multicolumn{1}{c|}{336}    &   - &  -    &  \multicolumn{1}{c|}{100}    &   \multicolumn{1}{c|}{381}                 &  -     & -    & 36.2     &  26.8    &  37   \\\hline
\begin{tabular}[c]{@{}l@{}}loop-\\else\end{tabular}   &  \multicolumn{1}{c|}{\textcolor{black}{128}}           &  -     &   - &  \multicolumn{1}{c|}{100}      &  \multicolumn{1}{c|}{\textcolor{black}{198}}                    & -      &  -     &  4.5   &  -    &   95.5    \\\hline
\begin{tabular}[c]{@{}l@{}}ass-m\\ul-tar\end{tabular}  & \multicolumn{1}{c|}{\textcolor{black}{174}}  &   50 &   2.5   &  \multicolumn{1}{c|}{47.5}    &          \multicolumn{1}{c|}{371}           & 16.2      &-      &1.6& -      &    82.2 \\\hline

\begin{tabular}[c]{@{}l@{}}star-\\in-call\end{tabular}  &   \multicolumn{1}{c|}{321}    & -   &  100  &   \multicolumn{1}{c|}{-}   &       \multicolumn{1}{c|}{\textcolor{black}{170}}              &	-         & 35.1     &  -    &  -    & 64.9 \\\hline
\begin{tabular}[c]{@{}l@{}}for-m\\ul-tar\end{tabular}  & \multicolumn{1}{c|}{356}     &  24.2    &    -&   \multicolumn{1}{c|}{75.8}    &     \multicolumn{1}{c|}{179}               &   67.6        &  -    &-      &  3.5    &  28.9 \\\hline
\end{tabular}
\vspace{-0.65cm}
\end{table}

Three categories (R1-R3) are shared by synthetic and real-project code, while two (R4, R5) are unique to real-project code which causes the inconsistencies between synthetic and real-project code.
For each idiom, we calculate the percentage of the examined idiomatic code instances whose performance is primarily affected by a root cause (Rx columns), resulting in either speedup or slowdown.
For example, for list-comprehension, 18.7\% of 310 examined idiomatic code are slower than non-idiomatic counterparts, which is primarily caused by R1, and the rest 81.3\% are faster, which is primarily caused by R2.

\begin{figure}[t]
  \centering
  
  \includegraphics[width=2.7in]{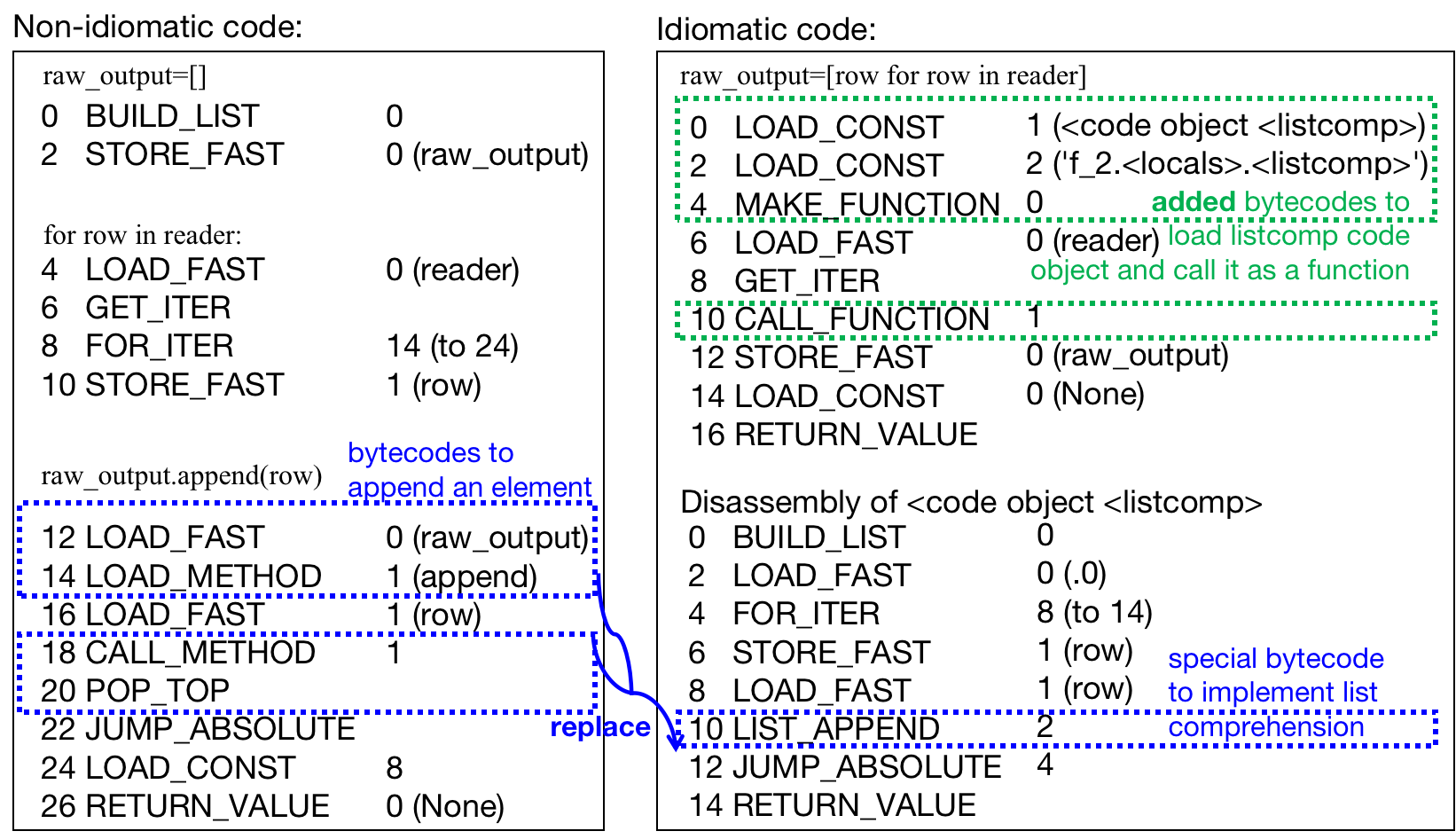}
\setlength{\abovecaptionskip}{-0.1cm}
    \caption{Bytecode Instructions of List Comprehension Code (Right) and Corresponding Non-Idiomatic Code (Left).  The examples for other idioms can be found in our \makebox[\linewidth]{\href{https://github.com/idiomaticrefactoring/PerformancePythonIdioms/blob/main/performance_python_idioms_supplementary_document.pdf}{supplementary document}} (see APPENDIX C.B)}
    
    \vspace{-7.5mm}
    \label{fig:bytecodelistcomp} 
\end{figure}

\noindent $\bullet$ \textbf{R1-Python idioms add new idiom-preparation bytecode which incurs execution overhead and slows down the overall performance.} 
The root cause is applicable to list/set/dict comprehension, assign-multi-targets and for-multi-targets. 
For list/set/dict comprehension, they need to execute additional preparation instructions (green box in Fig.~\ref{fig:bytecodelistcomp}) before performing iteration.
When there are no elements to append, the overhead of executing preparation instructions outweighs the time reduction by the special idiomatic instruction LIST\_APPEND (see R2), which results in the overall slowdown performance.

For assign-multi-targets, to assign each value to the corresponding target, it needs to additionally execute the BUILD\_TUPLE instruction to build a tuple and another UNPACK\_SEQUENCE instruction to unpack the sequences to put values onto the stack right-to-left. 
Therefore, when assign-multi-targets are only to assign variables, these additional instructions slowdowns the overall performance. 
The more assign nodes, the higher overhead by BUILD\_TUPLE and UNPACK\_SEQUENCE, the slower the performance.
Swapping variables and assigning constants are specially processed which can speedup the performance (see R2). 

For-multi-targets needs to execute unpacking instruction and storing unpacked values for later  use in the body of for statement. 
In real-project code, 90\% of for-multi-targets access part of items, so the idiom additionally needs a Starred node to store unused items. 
Besides, for-multi-targets usually accesses an item few times (2$\sim$6).
Therefore, 67.6\% of for-multi-targets slows down the performance in real-project code.

\noindent $\bullet$ \textbf{R2-Python idioms use specialized bytecode instructions to replace bytecodes of non-idiomatic codes, which could slows down or speed up overall performance.}
The root cause is applicable to list/set/dict-comprehension, assign-multiple-targets and chain-comparison. 
List-comprehension and set-comprehension have the same mechanism.
Take the list comprehension (blue box in Fig.~\ref{fig:bytecodelistcomp}) as an example.
The non-idiomatic code executes the LOAD\_FAST instruction to push data into the stack and then calls a function to append the element.
In contrast, the idiomatic code only needs to execute the LIST\_APPEND instruction to append the elements. 
Since the time of LIST\_APPEND is less than that of a series of instructions especially the expensive function call in the non-idiomatic code, the time for appending elements is shortened. 
As the number of added elements increases, the greater the time difference between executing the idiomatic and non-idiomatic instructions becomes.
When the \texttt{size} increases over a certain value, the speedup flattens because it essentially becomes the speedup ratio between the LIST\_APPEND and the non-idiomatic instructions.

The dict-comprehension executes the special instruction MAP\_ADD to load the dictionary iterable. Unlike the non-idiomatic code for list/set-comprehension, the non-idiomatic code for dict-comprehension does not need to execute function call.
As such, MAP\_ADD achieves basically the same thing as the instruction to store key-value pairs in non-idiomatic code. 
Therefore, the performance speedup of dict-comprehension is relatively smaller than that of list/set-comprehension. 

For assign-multi-targets, when swapping two or three variables, assign-multi-targets do not execute instructions to build and unpack a tuple, but use more cheaper instructions of ROT\_TWO and ROT\_THREE to directly swap variables, which speed up the performance. 

The chain-comparison replaces an instruction to load
a comparator of non-idiomatic code with the instructions to
create a reference to a comparator (DUP\_TOP) and reference shuffling (ROT\_THREE). 
Saving an extra reference to the comparator for the chain-comparison is not free, and cost more time than loading a local variable. 
Thus, if the comparator is a local variable, the performance decreases. 
Otherwise, the performance improves.

The star-in-func-call replaces instructions to load an element by index (BINARY\_SUBSCR) with instructions to build a slice object and unpack the slice object into arguments. When the star-in-func-call does not have the Subscript node, it does not execute instruction to build a slice object, which cost less time than the non-idiomatic code and results in speedup. 
When the star-in-func-call has the Subscript node, the execution time of instructions to build and unpack a slice object is more than that of one BINARY\_SUBSCR instruction.
So if the non-idiomatic code only has one Subcript node, the performance of star-in-func-call is slower. 
As the number of Subscript nodes of non-idiomatic code increases, the non-idiomatic code has to execute more and more BINARY\_SUBSCR instructions, which takes more time than the execution time of instructions to build a slice object and unpack the slice object.
In such cases, star-in-func-call results in speedup. 


\noindent $\bullet$ \textbf{R3-Python idioms remove bytecode instructions in non-idiomatic code which speeds up the performance.} 
This root cause is applicable to truth-value-test, loop-else, assign-multi-targets and for-multi-targets. 
The truth-value-test removes the instructions for loading the object from \texttt{EmptySet} and comparison. 
The loop-else removes the instructions for assigning the flagging variables and comparison with the flagging variables. 
If assign-multi-targets is to assign constant values, it removes the instructions to load constants. 
If assign-multi-targets is to swap variables, it removes the instructions to load and store temporary variables. 
The for-multi-targets remove multiple instructions to access an element by the index. 

\noindent $\bullet$ \textbf{R4-Non-Python built-in objects overloads Python built-in methods called by idioms, which may diminish the speedup, especially significantly affect the performance of the truth-value-test.}
This root cause is applicable to list/set/dict-comprehension, truth-value-test and for-multi-targets.
The relevant bytecode instructions are: iterating elements (FOR\_ITER) for list/set/dict-comprehension and for-multi-targets, comparison operation (COMPARE\_OP) and testing the truthiness (POP\_JUMP\_IF\_TRUE or POP\_JUMP\_IF\_FALSE) for truth-value-test.

FOR\_ITER calls \textsf{\_\_next\_\_} method to get the next item. 
If the object type is non-Python built-in type, its overloaded \textsf{\_\_next\_\_} can be slow, which may diminish the speedup of idiomatic instructions. 
For example, for the \href{https://github.com/kellyjonbrazil/jc/blob/e2f1b16cb9e920968c093bd1f371b191aa7107bf/jc/parsers/csv.py#L145-L146}{list-comprehension}: \textsf{[row for row in reader]}, the \textsf{reader} is a reader object from the CSV library. The overloaded \textsf{\_\_next\_\_} of \textsf{reader} cost more time than appending an element, so the bottle neck of the code is FOR\_ITER. Since FOR\_ITER of list comprehension and the corresponding non-idiomatic code is the same, the overall performance of idiomatic code is almost identical to the non-idiomatic code (1.008) no matter how many elements are added. 


The comparison operation may be overloaded by a library. Since such overloaded comparison operation in non-idiomatic code can be costly, replacing it with truth-value-test can significantly speed up the performance.
For example, the non-idiomatic code \href{https://github.com/nipy/nibabel/blob/225892af18d82b9a91ab63a9d072b894588b670e/nibabel/volumeutils.py#L940}{\textsf{if inter !==0}} is about 11 times slower than the truth-value-test \textsf{if inter} (the maximum speedup in Fig.~\ref{fig:performance_rq1}).
\textsf{inter} is an array object of the Numpy library, which overloads the comparison operation.
This overloaded comparison operation executes a series of dispatching, type conversion and wrapper object allocation.
The truth-value-test does not execute this comparison operation, which results in a significant speedup. 

The instruction to test the truthiness calls the \textsf{\_\_bool\_\_} method or \textsf{\_\_len\_\_} method if the object is not Boolean type. 
If the object is non-Python built-in data type, the implementation of \textsf{\_\_bool\_\_} or \textsf{\_\_len\_\_} may slow down the performance. 
For example, for the \href{https://github.com/searx/searx/blob/5b50d7455a719b145be5f90069026a60b480673d/searx/utils.py#L259}{\textsf{xpath\_results==[]}} code, the corresponding idiomatic code is \textsf{xpath\_results}. The \textsf{xpath\_results} is a \textsf{HtmlElement} object from the lxml library and implements the \textsf{\_\_bool\_\_} method inefficiently. 
So when executing the truth-value-test, it costs more time than the non-idiomatic code and slowdown the performance about 20 times (the slowdown outlier in Fig.~\ref{fig:performance_rq1}). 

\noindent $\bullet$ \textbf{R5-Real-project code involves more complex computations (i.e., the usage of non-Python built-in object, API calls, attribute access, compound expressions) which may diminish or change the impact of Python idioms.}
The root cause applies to all nine idioms as the primary root cause for the code instances ranging from 28.9\% to 95.5\%. 
R5 accounts for over 50\% for five idioms (list/dict-comprehension, loop-else, assign-multi-targets and star-in-func-call)

For the list/set/dict-comprehension, loop-else, truth-value-test, chain-comparison, assign-multi-targets and for-multi-targets, the variation of performance change is more likely to be smaller due to complex computation. 
For example, the \href{https://github.com/sympy/sympy/blob/429f5be2be5bdc5ee8d717ff857a3268b9c91e40/sympy/ntheory/residue_ntheory.py##L885-L887}{set comprehension} \textsf{\{pow(i,2,p) for i in range(p}\textsf{//2+1)\}} calls \textsf{pow} function that diminishes the speedup of set-comprehension ($\rho$=1.1). 
As another example, for \href{https://github.com/petertodd/python-bitcoinlib/blob/70ccd0863a5804b11d35e46176769a63aba28e3b/bitcoin/segwit_addr.py#L112}{the chain-comparison} \textsf{file\_size>self.max\_buffer\_size>0} and the non-idiomatic code \textsf{self.max\_buffer\_size>0 and file\_size>self.max\_buffer\_size}.
The chained comparator \textsf{self.max\_buffer\_size} is an attribute access which is slow, but creating and shuffling reference for the chained comparator cost less time than the time of loading \textsf{self.max\_buffer\_size} again. 
It leads to slight speedup ($\rho$=1.1).

The star-in-func-call with complex code is more likely to slow down the performance. Since the star-in-func-call requires additional execution of unpacking the object that accesses elements than the non-idiomatic code, if the object is non-Python built-in object, the unpacking instruction needs to construct a tuple for the object, which is costly. 
For example, for the \href{https://github.com/gugarosa/opytimizer/blob/f3e110cdc7fb8557d7e041737d3c9fd63f375c1e/opytimizer/visualization/surface.py#L41-L43}{non-idiomatic code} \textsf{func(points[0], points[1], points[2])}, the idiomatic code is \textsf{func(*points)}. 
If the \textsf{points} is Python built-in object (e.g., list and tuple), the idiom speeds up the performance. 
However, the \textsf{points} here is an array object from the Numpy library.
When unpacking the \textsf{points} for the star-in-func-call, it first construct a tuple for the \textsf{points} which cost more time than accessing elements by indices three times.
As a result, the idiomatic code becomes slower ($\rho$=0.7).


\noindent\fcolorbox{red}{white}{\begin{minipage}{8.6cm} \emph{The preparation instructions for Python idioms incur overheads, but Python idoms can shorten the execution time by using idiom-specialized instructions or performing special operations. As such, Python idioms can speedup the performance in the ``clean'' synthetic code by design. However, as the real-project code often involves library objects and complex computations, Python idioms in the real-project code generally do not produce as much performance impact as in the synthetic code or often result in slowdown.} \end{minipage}}

\section{Discussion}\label{sec:discuss}
\vspace{-0.1cm}
\subsection{Implications on Python Idiom Practices}

Although previous studies~\cite{alexandru2018usage,zhang2022making,leelaprute2022does} believe that Python idioms can improve performance, we find that developers are often confused and present contradictory opinions on the performance of nine Python idioms on Stack Overflow. 
Through our large-scale empirical study, we show that the outcomes of Python idioms on the synthetic code cannot be reliably transferred to the real-project code, and vice versa.
This is because the the designed speedup effects of Python idioms (albeit evident in the ``clean'' synthetic code) diminish in the real-project code involving complex non-Python built-in objects and computations.
Furthermore, even in the synthetic code, Python idioms sometimes slowdown the performance because the overhead of idiom preparation are not paid off by the reduced non-idiomatic operations.

When applying Python idioms, developers should consider code complexity, data size, variable scope and special code functions holistically.
Complex code (e.g., nested loop, many if-else checking, many unpacking targets) generally diminishes the effects of Python idioms.
Larger data size generally benefits more from list/set/dict-comprehension and for-multi-targets, but the speedup is capped by the time reduction ratio between the idiom-specialized processing and non-idiomatic processing.
Loading global variable takes more time than loading local variable.
Reducing the times of global variable loading (e.g., by the chain comparison) improves the performance, but the effect is much smaller for local variables.
Special code functions (e.g., truth-value-test against the values in EmptySet, swap variables or assign constants by assign-multi-targets, loop-else) benefit from the idioms as these idioms are specially designed to handle these code functions. 

When the real-project code is simple and does not involve non-Python built-in objects or complex computations (essentially similar to synthetic code), using Python idioms is generally beneficial.
However, when that is not the case, developers should be cautious as Python idioms may not bring in the designed speedup.
In real-project code, our results show that only truth-value-test may significantly speed up the performance, but chances are it may slow down the performance when it involves non-Python built-in objects that overload the truthiness test functions.
For-multi-targets suffers from the same slowdown symptom.
For other idioms, the differences between using or not using idioms are generally small.
As Python idioms squeeze many code tokens in a concise (sometimes obscure) form,
developers may be concerned about the readability of idiomatic code~\cite{zhang2022making}.
It may not be worth sacrificing the readability for some trivial performance speedup or even the risk of slowdown.  

\subsection{Threats to Validity}

The major \textbf{construct threat} is the performance variations due to non-determinism of code execution.
To mitigate this threat, we adopt mature performance measurement methods~\cite{georges2007statistically,laaber2019software,traini2021software}, and confirm the measurement reliability by a statistical sampling method~\cite{traini2021software,kalibera2020quantifying}.
One \textbf{internal threat} is the personal bias in analyzing the root causes of the performance changes. 
To reduce the bias, two authors with more than five years Python programming experience check the code instances independently, and then discuss to reach the consensus on the root causes. 
The other \textbf{internal threat} is the quality of our datasets. 
We set different parameters for code synthesis which well cover real code situations and use a large dataset of real-project code. 
To avoid human errors, we use a refactoring tool~\cite{zhang2022making} to automatically obtain idiomatic code from non-idiomatic code. 
For the \textbf{external threat}, our results are limited to the nine Python idioms and CPython 3.7 interpreter. 
Our research method can be extended to other idioms and interpreter versions.

\section{Related Work}\label{sec:relatedwork}
As Python idioms are highly valued by developers~\cite{programming_idioms,hettinger2013transforming,knupp2013writing,slatkin2019effective}, many studies mine Python idioms~\cite{alexandru2018usage,Zen_Your_Python,merchantepython,zhang2022making}. 
Some studies review popular Python books to create catalogues of pythonic idioms~\cite{alexandru2018usage,Zen_Your_Python,phan2020teddy}.
Many of these idioms are frequent code idioms also appearing in other languages.
Zhang et al.~\cite{zhang2022making} compare the language syntax of Python and Java to identify nine unique Python idioms not appearing in Java syntax, and develop an automatic code refactoring tool for nine Python idioms. 
Although some studies propose that Python idioms can improve the performance~\cite{alexandru2018usage,zhang2022making,leelaprute2022does}, there has been no large-scale empirical study on the performance impact of Python idioms. 
Our work fills in this gap.
Many studies~\cite{chen2019analyzing,luo2016mining,nguyen2012automated,nguyen2014industrial,trubiani2018exploiting,chen2020perfjit,Performance_Regression_huang,song2017performance,demeyer2005refactor} analyze source code to identify performance regression or improvement. 
Chen et al.~\cite{chen2020perfjit} extract metrics from code commits to predict whether commits cause performance regression. 
Traini et al.~\cite{traini2021software} extract 55 refactoring operations from the commits of 20 Java systems and execute the benchmarks before and after a commit. 
Leelaprute et al.~\cite{leelaprute2022does} analyzes the performance of some Python features (e.g., collections.defaultdict, lambda, generator expression) and two idioms (list/dict comprehension) with different input sizes. 
For list/dict comprehension, they find that as the number of added elements increases, the time difference between idiomatic code and non-idiomatic code becomes larger, which is consistent with our result.
However, since they only compare the time difference on one toy code example, they do not observe the wide distribution of performance differences of list/dict comprehension as our results show (see Fig.~\ref{fig:performance_rq1}), neither the discrepancies between synthetic and real-project code.
Furthermore, our study involves 7 more idioms and is done on two large scale datasets.

\section{Conclusion and Future Work}\label{sec:conclusion}

This paper presents a large-scale empirical study on the performance impact of nine Python idioms (list/set/dictionary comprehension, chain comparison, truth value test, loop else, assignment multiple targets, star in function calls, and for multiple targets) on both synthetic and real-project code.
We systematically present the wide performance-change distributions for different idioms with large-scale empirical evidence, and show that the performance outcomes of Python idioms in the synthetic code cannot be reliably transferred to the real-project code, and vice versa.
Our correlation and root-cause analysis show that these discrepancies are the outcomes of the joint force exerted by not only Python idiom design but also complex intrinsic and extrinsic code features.
Our results help developers make a holistic consideration of these interweaving factors when using Python idioms and develop a realistic understanding of the pros and cons of Python idioms.

\bibliographystyle{IEEE-Reference-Format}
\balance
\bibliography{debug}

\end{document}